\documentclass[twocolumn,showpacs,amsmath,amssymb,pre,aps]{revtex4} %%  4-pages !!
\usepackage{graphicx}% Include figure files
\usepackage{dcolumn}% Align table columns on decimal point
\usepackage{bm}% bold math
\usepackage{color}% color text
\usepackage{amsmath}

\renewcommand{\vec}[1]{\mathbf{#1}}

\newif\ifgraph

\graphtrue             %
\begin{document}

\title{Diffusion Transients in Motility-Induced Phase Separation}

\author{Shubhadip Nayak$^{1}$, Poulami Bag$^{1}$, Pulak K. Ghosh$^{1}$}
\email{pulak.chem@presiuniv.ac.in}
\affiliation{$^{1}$ Department of Chemistry, Presidency University, Kolkata 700073, India}
\author{Yunyun Li$^{2}$}
\email{yunyunli@tongji.edu.cn}
\author{Yuxin Zhou$^{2}$, Qingqing Yin$^{2}$, Fabio Marchesoni$^{2,3}$}
 \affiliation{$^{2}$ MOE Key Laboratory of Advanced Mico-Structured Materials, School of Physics Science and Engineering, Tongji University, Shanghai 200092, China}
 \affiliation{$^{3}$ Dipartimento di Fisica, Universit\`{a} di Camerino, I-62032 Camerino, Italy}

% \email{fabio.marchesoni@pg.infn.it}
\author{Franco Nori$^{4,5}$}
\affiliation{$^{4}$ Theoretical Quantum Physics Laboratory, Quantum Computing Center, RIKEN, Wako-shi, Saitama 351-0198, Japan}
\affiliation{$^{5}$ Physics Department, University of Michigan, Ann Arbor, Michigan 48109-1040, USA}
\date{\today}

\begin{abstract}
We numerically investigate normal diffusion in a two-dimensional athermal suspension of active particles undergoing motility-induced phase separation. The particles are modeled as achiral Janus disks with fixed self-propulsion speed and weakly fluctuating orientation. When plotted versus the overall suspension packing fraction, the relevant diffusion constant traces a hysteresis loop with sharp jumps in correspondence with the binodal and spinodal of the gaseous phase. No hysteresis loop is observed between the spinodal and binodal of the dense phase, as they appear to overlap. Moreover, even under steady-state phase separation, the particle displacement distributions exhibit non-Gaussian normal diffusion with transient fat (thin) tails in the presence (absence) of phase separation.

\end{abstract}
\maketitle

\section{Introduction} The normal diffusion of an ideal massless Brownian
particle is usually associated with the Gaussian distribution of its spatial
displacements. However, there are no fundamental
reasons why the diffusion of a physical Brownian tracer should be of the
Fickian type \cite{Granick1,Granick2,Sung2,Granick3}. For instance, displacement
distributions in real biophysical systems appear to retain prominent exponential tails,
even after the tracer has attained the condition of normal diffusion. Such an
effect, often termed non-Gaussian normal diffusion (NGND), disappears only
for exceedingly long observation times (possibly inaccessible to real
experiments \cite{Granick1}), when the tracer's displacement distribution
eventually turns Gaussian, as dictated by the central limit theorem
\cite{Gardiner}. Persistent diffusive transients of this type have been
detected in experimental and numerical setups
\cite{Sung1,Tong,Jain1,Bhatta,Cherstvy}. The signature of NGND, along with a Non-Gaussian velocity distribution, has been previously reported in systems with spatial heterogeneity\cite{Lemaitre,Khadem,Nanoscale}. 
The current interpretation of such
diverse NGND manifestations as transient effects, postulates the existence of one or
more slowly fluctuating processes affecting composition, geometry and dynamics
of the tracer’s environment \cite{Slater,Metzler2,Jain2,Tyagi,Sokolov,NGND-our,FoP}.

In this paper we report conspicuous manifestation of NGND in a one-component suspension of
identical micro-swimmers undergoing phase separation. The term
micro-swimmers refers to either motile micro-organisms, like bacteria, or
their synthetic counterparts, typically two-faced colloidal particles [for
this reason called Janus particles (JP)], both capable of self-propulsion
  under non-equilibrium conditions
\cite{Granick,Muller,Marchetti,Gompper}.  Artificial swimmers are a topic
of current research as these can be designed and operated as micro-robots for
specific applications \cite{Sen,Wang}.

When investigated collectively, a suspension of active particles may
undergo phase separation even in the absence of cohesive forces
\cite{Fily,Redner1,Redner2}. The ensuing motility-induced phase separation
(MIPS) is arguably the simplest non-trivial collective feature that
distinguishes active from passive particles \cite{Tailleur}. MIPS involves
the coexistence of two active phases of different densities, similarly to
what happens in a binary fluid mixture below its critical temperature. It
occurs as a combined effect of steric interactions and self-propulsion, even
in the absence of pair alignment, interactions with solid substrates or
thermal fluctuations \cite{Fily,Redner1}. Experimental evidence of MIPS has
been obtained both in biological and synthetic systems,
%\cite{Bechinger}
despite numerous technical difficulties \cite{Tailleur}.

Inspired by its similarity with equilibrium phase decomposition, much
effort has been made to describe MIPS in terms of
a non-equilibrium phase transition theory \cite{Omar_Brady,Nardini,Tailleur}.
In particular, numerical observations \cite{Redner2} and field theory arguments
\cite{Tailleur2} confirm that, the phase diagram of an active suspension
exhibits distinct binodal and spinodal lines: in the binodal region enclosed
between them, the suspension is in a metastable homogeneous phase, which
undergoes slow phase separation through delayed nucleation and fast growth, while
near and inside the spinodal region it separates by fast coarsening.

Among the quantitative tools employed to numerically characterize MIPS,
diffusivity offers arguably the most direct access to the microscopic
dynamics underlying phase separation. The asymptotic diffusion constant has
been computed as an overall indicator of both gas-liquid \cite{Fily} and
liquid-solid separation \cite{Loewen1}. Diffusivity was utilized also
to analyze the inner structure
of the separating clusters \cite{Redner1}.

In this paper we show that
diffusion in an athermal active suspension under MIPS may provide a more
predictive tool than previously reported. To avoid more complex phase
diagrams \cite{3D1,3D2}, we restrict this report to a two-dimensional (2D)
suspension of active hard disks. Such disks undergo normal diffusion
no matter what the suspension phase. Upon
increasing the suspension packing fraction with uniform initial particle
distribution, the diffusion constant exhibits a sharp drop, which we interpret
as the gaseous phase spinodal. However, slowly ramping up and down the
overall packing fraction, produces a robust hysteresis loop delimited by
the binodal and spinodal of the gaseous phase. Vice versa, within our numerical accuracy, the
binodal region of the dense phase appears to collapse, so that no
hysteretic diffusion loop was observed. Moreover, in the presence of MIPS, 
the corresponding particle displacement distributions are
leptokurtic for extended time transients (i.e., tend to zero slower than a Gaussian function),
a clear-cut NGND manifestation. 

\section{Model} We simulated a two dimensional suspension of $N$ identical achiral active JP's
modeled as disks of radius $r_0$ and constant self-propulsion speed, $v_0$,
in a square box of size $L$ with periodic boundary conditions. The dynamics of a
single JP of coordinates ${\vec r}=(x,y)$ obeys the simple Langevin equations,
\begin{eqnarray} \label{LE}
\dot {\vec r}=  {\vec v}_0, \;\; \dot \theta =\sqrt{D_\theta}~\xi_\theta (t).
\end{eqnarray}
Here the orientation of the self-propulsion vector, ${\vec v}_0=v_0(\cos
\theta, \sin \theta)$, measured with respect to the longitudinal $x$-axis,
fluctuates subjected to the stationary, delta-correlated noise source
$\xi_\theta (t)$, with $\langle \xi_\theta(t)\xi_\theta(0)\rangle = 2 \delta
(t)$. Following Ref. \cite{Fily}, the suspension is assumed to be athermal,
that is, we neglect thermal fluctuations against the angular
noise intrinsic to the self-propulsion mechanism \cite{ourPRL,ourJPCL}. The
reciprocal of $D_\theta$ defines the correlation time, $\tau_\theta$, and the
persistence length $l_\theta=v_0/D_\theta$ of a free self-propelled
JP. For $t\gg \tau_\theta$, a free JP would undergo normal diffusion
with diffusion constant $D_s=v_0^2/2D_\theta$, but non-Gaussian statistics.

\begin{figure}[tp]
\centering \includegraphics[width=8.0cm]{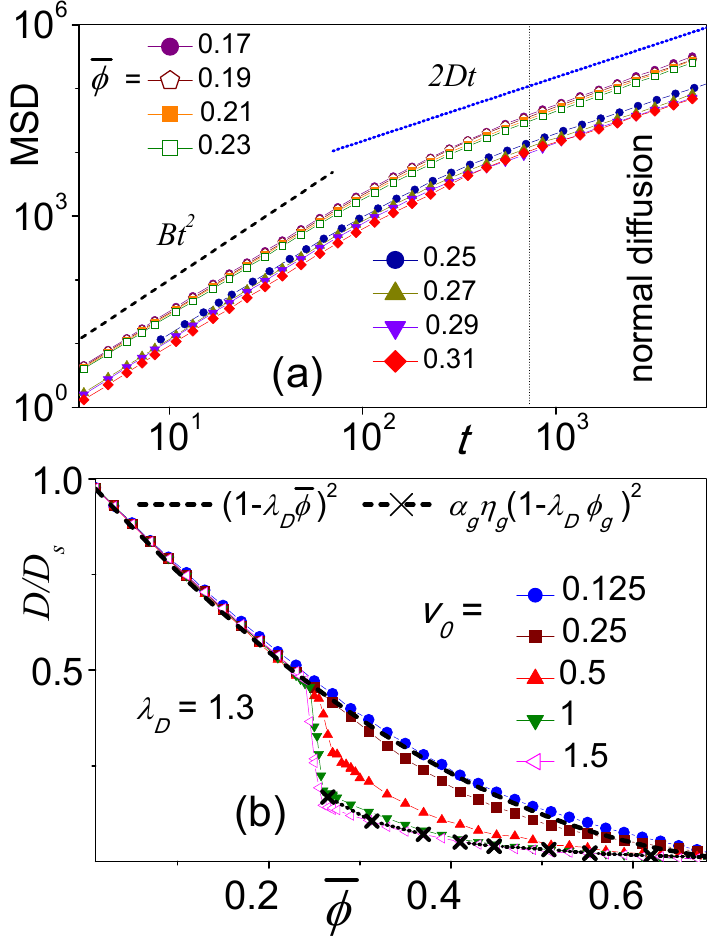}
\centering \includegraphics[width=7.8cm]{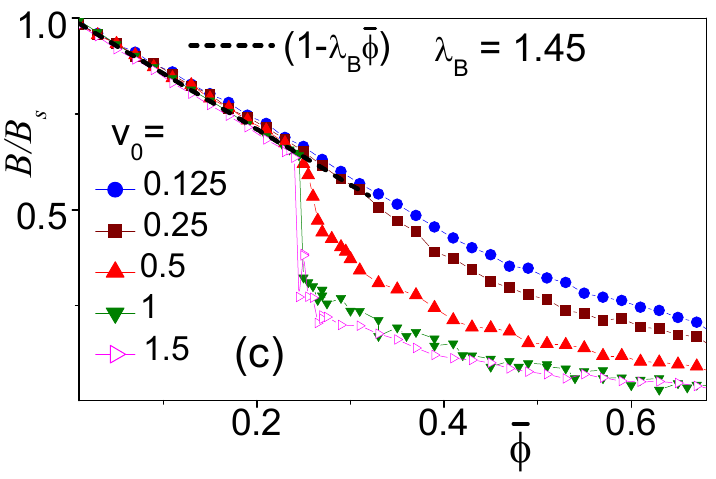}
\caption{Particle mean-square displacement (MSD) in a suspensions of $N$ active JP's of radius $r_0=1$ and persistence time $\tau_\theta=100$; $N=9,120$; Packing fraction, $\bar \phi$, was varied by tuning simulation box size at fixed $N$. If not stated otherwise, the suspension was initially randomly uniform. (a) MSD, $\langle \Delta x^2(t) \rangle$, vs. $t$ for $v_0=1$ and different $\bar \phi$. The short-time ballistic and the asymptotic diffusive branches are fitted respectively by quadratic, $Bt^2$, and linear, $2Dt$, functions. The fitting parameters, $D$ (in units of $D_s=v_0^2/2D_\theta$), are plotted in (b) vs. $\bar \phi$. The fitting parameter, $B$ (in units of $B_s=v_0^2/2$), vs. $\bar \phi$  with different $v_0$ (see legend) are plotted in (c). The dashed lines in (b) and (c) depict the $\bar \phi$ dependence of $D/D_s$ and $B/B_s$ (for $v_0=1$) are fitted by the functions in the legends. Crosses in the panle (b) display Eq.~(\ref{D_mips})[see text for details]. \label{F1}}
\end{figure}

At short distances the disks repel each other via the %Weeks-Chandler-Andersen (WCA)
pair potential \cite{WCA},
\begin{eqnarray}\label{WCA}
V_{ij} &=& 4\epsilon [({\sigma}/{r_{ij}})^{12} -({\sigma}/{r_{ij}})^{6} +1/4], \;\; {\rm if}\;\;  r_{ij} \leq r_m \nonumber \\
  &=& 0 \;\; {\rm otherwise},
\end{eqnarray}
where $i,j=1, \dots N$ are the pair labels, $r_m=2^{1/6}\sigma$,
$\epsilon=1$, and $\sigma
= 2r_0$ represents the ``nominal'' disk diameter. The steric interactions of Eq.
(\ref{WCA}) are not corrected for hydrodynamic effects \cite{Marchetti,PNAS}.
To save computer time, the suspension packing fraction, $\bar \phi=\pi r_0^2 N/L^2$, was varied by
changing the box size, while keeping the number and radius of the disks fixed.
The stochastic differential Eqs. (\ref{LE}) were integrated numerically by
means of a standard Euler-Maruyama scheme \cite{Kloeden}.

As illustrated in Sec.~III and IV 
%( also in SM \cite{SM}),
 the MIPS phenomenon was readily reproduced at fixed $v_0$ by increasing $\bar \phi$.
Upon approaching the MIPS onset at $\bar \phi=\bar \phi_*$, short-lived aggregates
in the homogeneous phase anticipate the separation into a
gaseous (or dilute) and a dense phase: for $\bar \phi<0.5$
($\bar \phi>0.5$) a single large cluster (cavity) forms in the simulation box.
Most notably, tagged particles diffuse homogeneously across the simulation box also
in the presence of phase separation, i.e., for $\bar \phi > \bar \phi_*$, no matter
what the cluster size.

A JP of speed $v_0$ and persistence time $\tau_\theta$ is characterized
by a mean ballistic path, or persistence length, $l_\theta=v_0\tau_\theta$.
This dynamical length should be compared with the other characteristic
length scales in the homogeneous phase. As known from the
classical kinetic gas theory, they are (i) the average particle distance,
$l_L=\sqrt{L^2/N}$, and (ii) the mean-free path between pair collisions,
$l_c=L^2/N\sigma$. We ignore here the particle diameter, $\sigma=2r_0$, as
we kept it comparatively small throughout our numerical investigation.
When expressed in terms of the overall suspension packing
fraction, $\bar \phi$, the lengths above read respectively, $$l_L=\sqrt{\pi r_0^2/\bar\phi}, \;\; {\rm and} \; \; l_c=\pi r_0/2\bar \phi.$$
This implies that, if we keep $\bar
\phi$ fixed and vary the particle number, $N$, the ratios $l_L/l_\theta$ and
$l_c/l_\theta$ do not change.

Being MIPS a collisional mechanism \cite{Brito}, we limited our investigation
to suspension densities with $l_c\geqslant l_L$, that is on $\bar \phi$
values not exceeding the close-collision packing fraction
$\phi_{\rm cc} = \pi/4$, an upper bound slightly smaller than the close
packing fraction, $\phi_{\rm cp}=2\sqrt{3}/\pi$, often mentioned in the
literature. 
\begin{figure}[tp]
\centering \includegraphics[width=7.5cm]{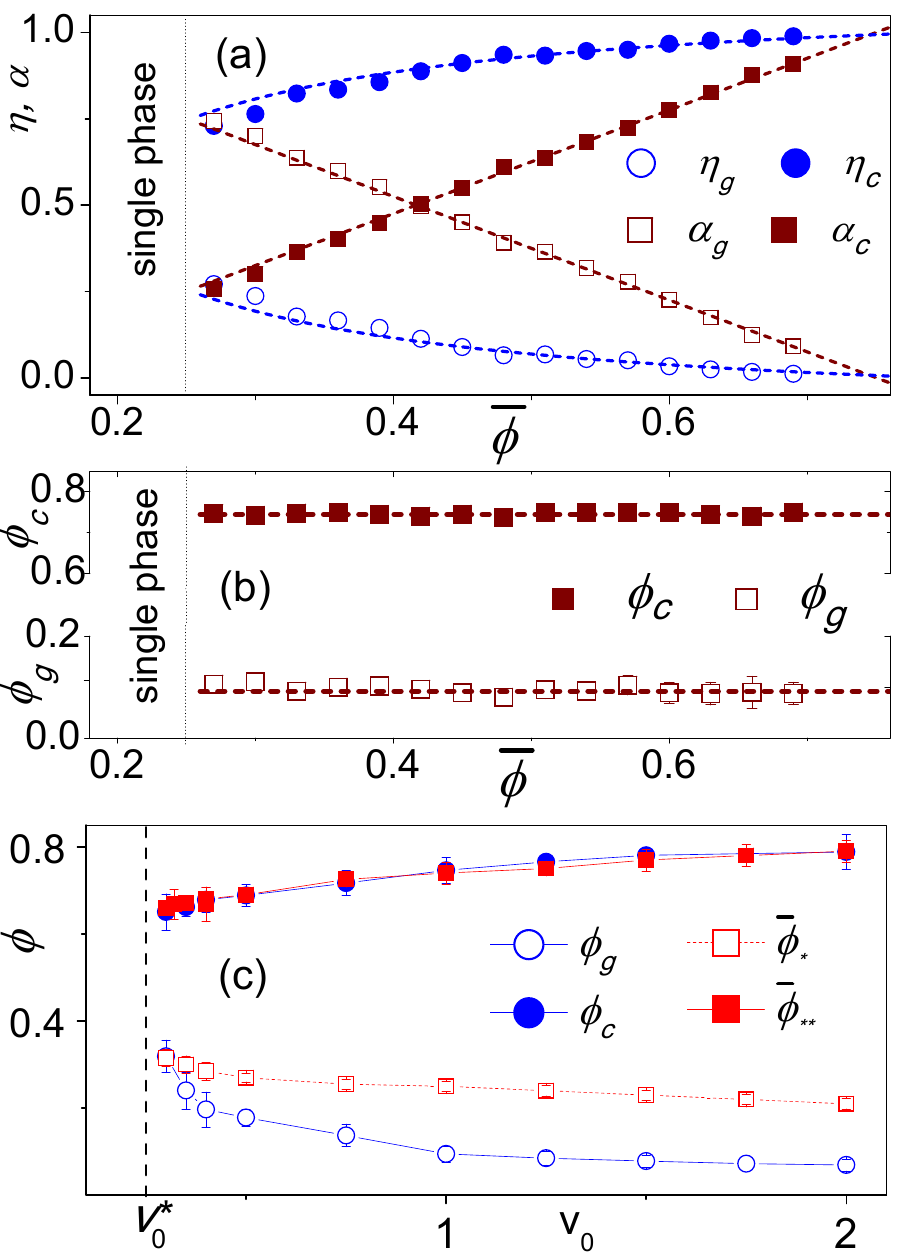}
\centering \includegraphics[width=7.1cm]{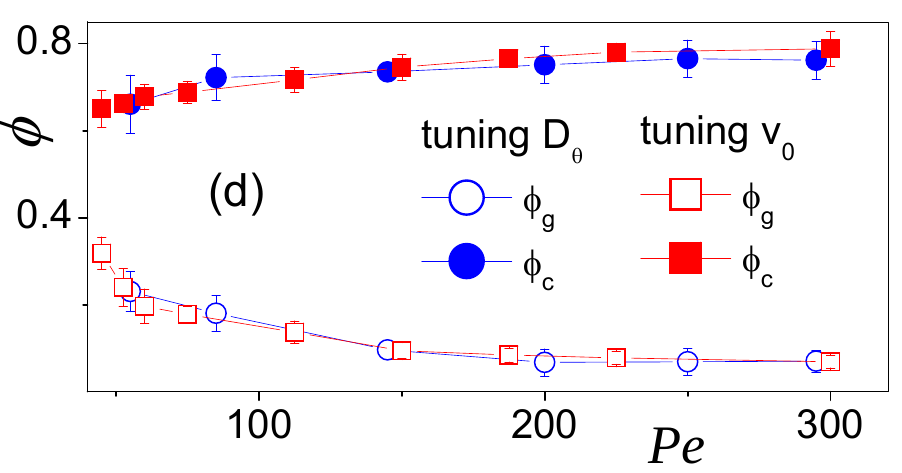}
\caption{Gaseous and dense phases (denoted respectively by the indices $i=g,c$): $\bar \phi$ dependence of the phase (a) number, $\eta_i$, and volume, $\alpha_i$, fractions and (b) the phase packing fractions, $\phi_i$, for $N=10^4$, $v_0=1$, $\tau_\theta=100$, and random uniform initial conditions. As a consistency test, we fitted the data for $\alpha_i$ and $\phi_i$ with straight lines [dashed lines respectively in (a) and (b)] and made use of the identity $\phi_i=\bar \phi \eta_i/\alpha_i$ (see text) to reproduce the $\bar \phi$ dependence of $\eta_i$ in (a). Fitting functions: $\alpha_c=1.5\bar \phi-0.14$, $\phi_c=0.75$, and $\phi_g=0.09$. Recall that $\alpha_g+\alpha_c=1$ and $\eta_g+\eta_c=1$.
(c) Cluster, $\phi_c$, gaseous phase, $\phi_g$, and MIPS onset, $\bar \phi_*$ and $\bar \phi_{**}$, packing fractions vs. $v_0$ for $\bar \phi=0.45$. For $v_0<v_0^*$ MIPS never occurs, while for $v_0>v_0^*$ our data suggest that $\phi^{\rm (s)}_g +\phi^{\rm (s)}_c=1$. (d) Gaseous and cluster phase binodal, $\phi_g$ and $\phi_c$, vs. $Pe=3v_0/(2r_0D_\theta)$ for $N=9,120$, $v_0=1$ (blue symbols) and $D_\theta=10^{-2}$ (red symbols).
\label{F2}}
\end{figure}
\section{Particle self-diffusion and motility-induced phase separation} To characterize the particle diffusivity in
the suspension, we monitored two quantities: (i) the particle mean-square
displacement (MSD), say, in the $x$ direction, $\langle \Delta x^2(t)
\rangle$, with $\Delta x(t)=x(t)-x(0)$. In view of the established ergodic character of
the diffusive process, ensemble averages,
$\langle \dots \rangle$, were taken over all $N$ suspension particles
\cite{PNAS}; (ii) the probability density function (pdf) of the particle
displacement $\Delta x$ at time $t$, $P(\Delta x,t)$. Both quantities were
computed after the suspension had reached an apparently steady-state
configuration. Note that $P(\Delta x,t)$ displays a small periodic oscillatory pattern due to the periodic boundary conditions applied in our simulation. However, this effect does not affect the findings reported in this paper. %(see Sec. II of SM\cite{SM}).

Our numerical data [Fig.~\ref{F1}(a)] clearly show that for sufficiently
large observation times, typically $t\gg \tau_\theta$, the MSD grows
according to the Stokes-Einstein law, $\langle \Delta x^2(t) \rangle = 2Dt$,
which defines the particle self-diffusion constant in the suspension, $D$. This
constant is a function of $\bar \phi$: A sudden %(but seemingly continuous)
diffusivity drop marks phase separation. This MIPS signature is sharp enough
to determine $\bar \phi_*$ as a function of $v_0$. As apparent in Figs.
\ref{F1}(b,c) and \ref{F2}(c,d), no MIPS occurs for $v_0$ below a critical value, $v_0^*
\simeq 0.25$, while for $v_0>v_0^*$ the dependence of $\bar \phi_*$
on $v_0$ is rather weak. Similarly, Fig.~\ref{F2}(d) shows that for a given $v_0$, there exists an upper bound for the angular diffusion constant $D_\theta$, above which MIPS does not occur. %For more details, see Sec IV of SM~\cite{SM}.    

In the homogeneous phase, the fitting ballistic, $B$, and diffusion, $D$,
constants of Fig.~\ref{F1}(a) appear to slowly decrease with increasing
$\bar \phi$ up to the MIPS onset, $\bar \phi=\bar \phi_*$. Standard stoichiometric arguments suggest polynomial fitting laws,
\begin{eqnarray}\label{D_gas} 
D=D_s(1-\lambda_D \bar \phi)^2, \;\; {\rm and} \;\; B=B_s(1-\lambda_B \bar \phi), 
\end{eqnarray}
with $D_s=v_0^2 \tau_\theta/2$ and $B_s=v_0^2/2$.
Both fitting parameters $\lambda_B$ [in Fig. \ref{F1} (c)] and $\lambda_D$
[in Fig. \ref{F1} (b)]  are larger than the
reciprocal of the close-collisional packing fraction, $\phi_{\rm cp}=\pi/4$,
introduced in Sec. II, and $\lambda_B > \lambda_D$. We attribute this behavior to
the residual softness of the repulsive WCA potential. Similar to the diffusivity curves, MIPS drops of the
curves  $B(\bar \phi)$ are apparent, in quantitative agreement with the existence of a critical value $v_0^*$, below which MIPS is ruled out [see Fig.~\ref{F2}(c,d)].  These findings are consistent with the conventional analysis of the stationary local packing fraction distribution. %(see Sec.III of SM\cite{SM}).

To analyze the tails of $D(\bar \phi)$ for $\bar \phi > \bar \phi_*$, we had
recourse to the two-phase characterization of Figs. \ref{F2}(a,b) (also for
$v_0=1$). After exceedingly long simulation runs, $t= 10^5$, the dense and dilute
phases of the suspension appear to be well separated. We computed the volume, $\alpha_i$, and number,
$\eta_i$, fractions of both phases and the resulting phase packing fractions,
$\phi_i$ ($i=g,c$ denoting, respectively, the gaseous and the dense phase). To
this purpose we first computed the corresponding phase densities $\rho_i$, by
selecting rectangular regions (as large as possible) within either phases and
then counting particles in there. This procedure was repeated $10$ times for
different ``trajectories'', namely initial configurations and 
random number sequences. This way we estimated the phase mean densities as
well as their standard deviations (under the simplifying assumption that both phases
were homogeneous). Finally, we computed the phase areas,
$A_i$, by imposing the two normalization conditions
$\rho_c  A_c + \rho_g  A_g = N$ and $A_c +  A_g = L^2$.

By definition, $\phi_i=\bar \phi \eta_i/\alpha_i$, as numerically checked in
Fig. \ref{F2}(a). The densities of the two phases are confirmed to be
independent of $\bar \phi$ \cite{Redner1}. On neglecting the contribution
from the particles trapped in the cluster, the self-diffusion constant for
$\bar \phi > \bar \phi_*$ can be approximated to,
\begin{eqnarray}\label{D_mips}
 D_{\rm MIPS}(\bar\phi)=\alpha_g \eta_g D_s(1-\lambda_D \phi_g)^2.
\end{eqnarray}
 Here, we made use of the
fact that the gaseous phase represents a fraction $\eta_g$ of the suspension
and behaves as a homogeneous phase with low packing fraction, $\phi_g$, and
fractional volume $\alpha_g$. A comparison with the actual $D$ data for
$v_0=1$ is displayed in Fig. \ref{F1}(b).

\begin{figure}[tp]
\centering \includegraphics[width=8.0cm]{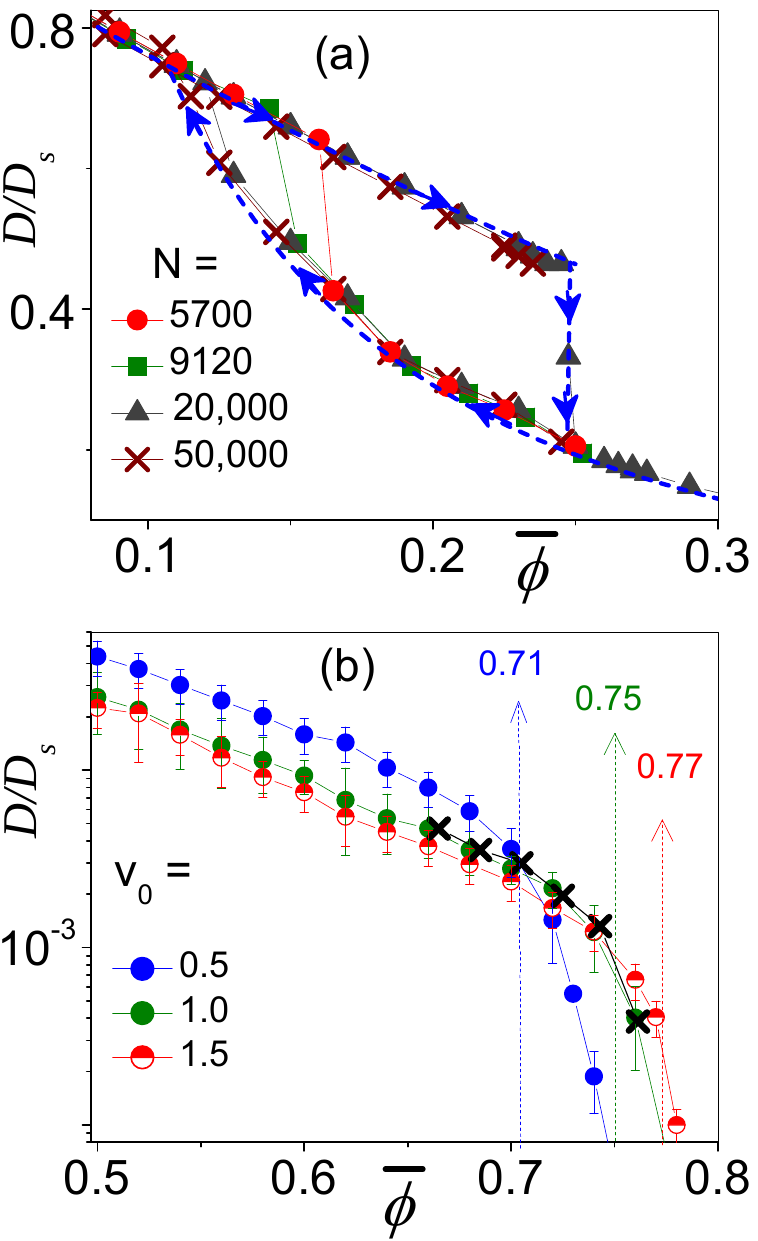}
%\centering \includegraphics[width=7.6cm]{Fig2d.pdf}
\caption{(a) Hysteresis loop obtained by slowly ramping $\bar \phi$ up and down across the gaseous binodal-spinodal range for $v_0=1$ with different $N$ (see legends). The reference hysteresis loop (dashed blue curve) has been closed by extending the lower fitting curve, $D_{\rm MIPS}(\bar \phi)$, in Fig.1(b) down to $\bar \phi=\phi_g$. (b) $D$ vs. $\bar \phi$ in the upper binodal region for different $v_0$ and $N = 20,000$. For $v_0=1$, $D$ was determined like in (a), by ramping $\bar \phi$ up (dots) and down (crosses); no hysteresis loop was detected, as the upper binodal region appear to collapse.
\label{F3}}
\end{figure}

\section{Hysteresis Loops and binodal points}

In order to clarify the meaning of $\bar \phi_*$, we notice that for
$\bar \phi >\bar \phi_*$ the cluster volume fraction, $\alpha_c$, grows
linearly with the overall packing fraction. Our simulation results for $\alpha_c (\bar \phi)$ fit well (within numerical error) with the following empirical relation, 
\begin{eqnarray}\label{alpha_c}
\alpha_c(\bar \phi)=(\bar \phi-\phi_g)/(\phi_c-\phi_g).
\end{eqnarray}
Where $\phi_c$($\phi_g$) is the packing fraction of a dense (dilute) one-phase suspension,
i.e., for $\alpha_c=1$ [$\alpha_c=0$, out of range in Fig. \ref{F2}(a)]. This
simple graphical construction allows one to quickly determine both binodal
points, $\phi_g$ and $\phi_c$, at fixed $v_0$, in good agreement with the
simulation of Fig.~\ref{F2}(b). Their dependence on $v_0$, displayed in Fig. \ref{F2}(c) for $\bar \phi=0.45$, qualitatively agrees with the simulation results of
Refs. \cite{Redner1,Redner2}. Note that we never increased $v_0$ large enough
to explore either the full phase diagram \cite{full_phase,Krauth} and/or
the re-entrant MIPS \cite{Hou}. Furthermore, the spinodal curves displayed in Fig. 2(c) remain unchanged as the system size increases up to $N = 50,000$. For this large system size, we conducted simulations with a duration of $t = 20,000$. However, these results may vary for significantly larger system sizes and longer simulation runs, which lie  beyond our computational capabilities. 

\subsection{The gaseous phase spinodal and binodal}

In contrast with Refs. \cite{Fily,Redner1}, by starting with a
uniform particle distribution we never observed MIPS in the range $\bar \phi
\in [\phi_g, \bar \phi_*]$, no matter what the (accessible) running time. The
outcome changed when we slowly increased (decreased) $\bar \phi$ over time.
We did so by keeping $N$ fixed and decreasing (increasing) $L$ stepwise after
a fixed long running time $\Delta t$ (typically $\Delta t=5\times 10^4$).
Upon varying $L$, we rescaled the suspension configuration accordingly. This
produced the hysteresis loops of Fig. \ref{F3}(a), which, for large $N$, approach the ideal loop obtained by
connecting the fitting functions of $D$ vs $\bar \phi$ in Fig. \ref{F1}(b) [also see Eqs.(\ref{D_gas}-\ref{D_mips})].
On increasing $\bar \phi$, MIPS occurs, as anticipated above, at $\bar
\phi_*$ (signaled by a $D$ drop), but upon decreasing $\bar \phi$, it only
disappears for $\bar \phi \gtrsim \phi_g$ (signaled by a fast $D$ rise).

The reference or ideal hysteresis loop in Fig. \ref{F3}(a) [blue dashed lines]  was obtained by
extending the $\bar \phi$-function $D_{\rm MIPS}$ [see Eq.~(\ref{D_mips})]  to $\bar \phi$ values
below the MIPS threshold, $\bar \phi_*$. On making use of  Eq.~(\ref{alpha_c}), the MIPS branch of
the loop can be written as,
\begin{eqnarray}
\frac{D_{\rm MIPS}}{D_s}=\left(\frac{\phi_c-\bar
\phi}{\phi_c-\phi_g}\right)^2 \frac{\phi_g}{\bar \phi}(1-\lambda_D \phi_g)^2,
\end{eqnarray}
hence $D_{\rm MIPS}(\phi_g)=D(\phi_c)$.

 To check robustness of the hysteretic effect toward translational noises, we simulate Eq.~(1) after adding a 2D translational Gaussian noise term with strength $D_0$,
${\bm \xi}(t)=(\xi_x(t), \xi_y(t))$ with $\langle \xi_i(t)\rangle=0$ and
$\langle \xi_i(t) \xi_j(0)\rangle=2D_0\delta(t)$ for $i,j=x,y$,
to Eq.~(1). The hysteresis loop of Fig. \ref{F3}(a) there turned
out to be quite robust; indeed, it appeared to vanish only for $D_0$ of the order of $D_s$.
Vice versa, its area may be quite sensitive to the suspension size, $N$. 

Recall that our hysteresis protocol $\bar \phi$ was
increased/decreased stepwise at regular time intervals, $\Delta t$.  Of
course, we cannot rule out the possibility that the resulting hysteresis loop
shrinks and finally disappears for exceedingly large $\Delta t$ (in any case,
well beyond our computing capabilities). Similar remarks apply to even
simpler dynamically bistable systems, like the motility of a weakly damped,
driven Brownian particle confined to a one dimensional washboard potential
\cite{Risken}. For the suspension of Fig. \ref{F3}(a), we repeatedly looped
$\bar \phi$ in the range $(0.05, 0.30)$, that is across the relevant binodal and spinodal of the gaseous phase.

As we verified that the hysteretic effect is robust toward translational noises, it has also been noticed that hysteresis loops become sharper upon increasing the suspension size, $N$, and the observation time, $t$. In conclusion, accurate data for the $D(\phi)$ hysteresis loop suffice to self-consistently
characterize the gaseous binodal region at fixed $v_0$. Further, the persistence of uniformly distributed short-time aggregates
in the suspensions with  $\bar \phi \lesssim \bar \phi_*$, suggests to interpret $\bar \phi_*$ as the gaseous phase spinodal \cite{Redner1,Redner2},
$\bar \phi_*=\phi^{\rm (s)}_g$.

\subsection{The cluster spinodal and binodal}
 A similar approach was adopted by
simulating initially homogeneous, dense suspensions and decreasing
$\bar \phi$ below $\phi_c$: at a sufficiently low value of the overall
packing fraction, $\bar \phi=\bar \phi_{**}<\phi_c$, the dense suspension
developed coalescing gaseous bubbles. Therefore, the curve $\bar \phi_{**}$ versus $v_0$ displayed
in Fig. \ref{F2}(c) is our best estimate of the cluster spinodal, $\phi^{\rm
(s)}_c$. As illustrated in Fig. \ref{F3}(b),
at $\bar \phi=\phi^{\rm (s)}_c$, the curves $D$ versus $\bar \phi$
do exhibit a second drop,  though not as sharp as at $\bar \phi=\phi^{\rm (s)}_g$, but
no hysteretic loop. In fact,
cluster binodal and spinodal curves run so close to one another that we could
hardly separate them; the upper binodal region appears to collapse 
[see Ref. \cite{Tailleur2} for an analytical treatment]. Remarkably
enough, our numerical data suggest that $\phi^{\rm (s)}_g +\phi^{\rm (s)}_c=1$.
As $v_0$ approaches $v_0^*$ (from above), both upper and lower pairs of binodal
and spinodal curves overlap.

\begin{figure}[tp]
\centering 
\includegraphics[width=7.5cm]{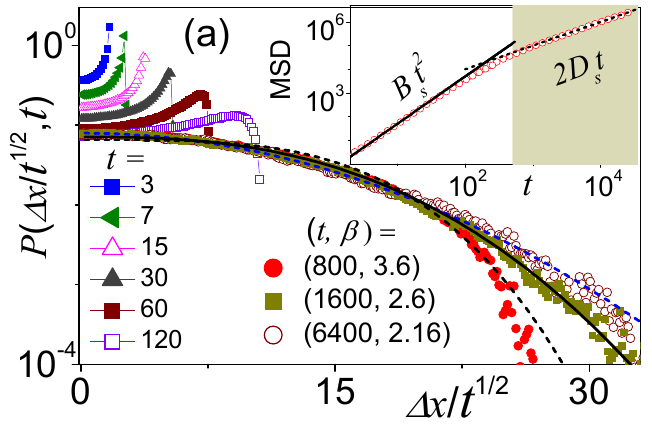}
\includegraphics[width=7.5cm]{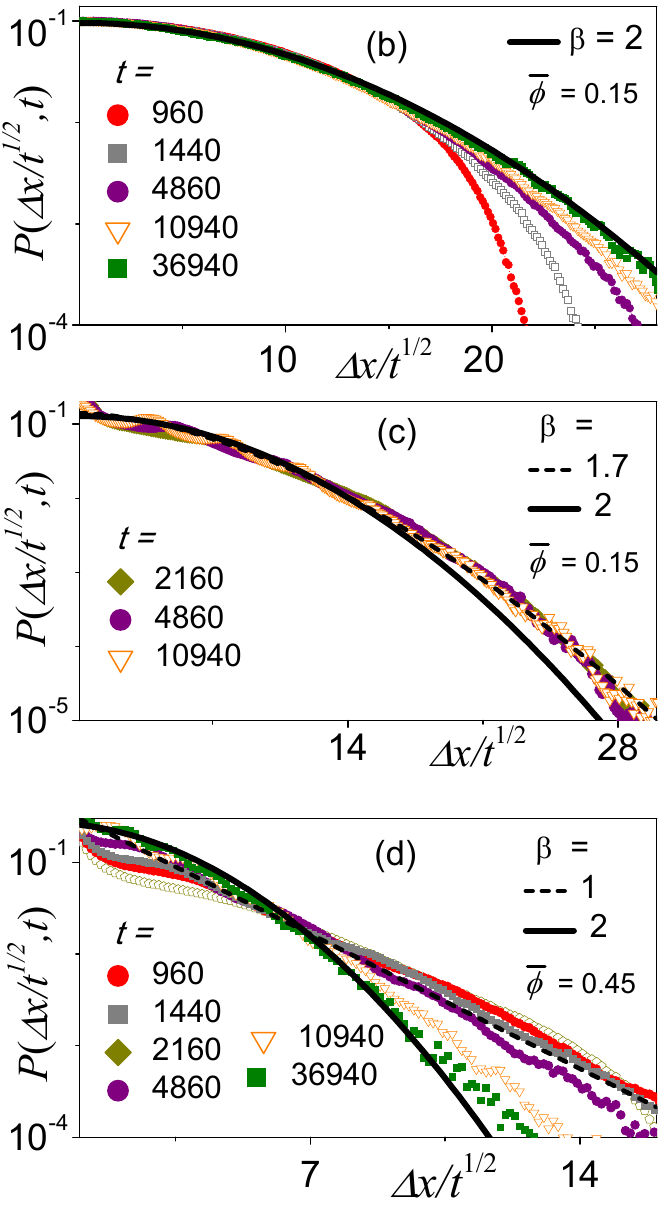}
\caption{Non-Gaussian normal diffusion as a MIPS signature. (a) Particle diffusion in an active suspension of non-interacting JP's, $\epsilon=0$, at different times, $t$ (see legend). The transient (platykurtic) fitting curves, $P(\Delta x/\sqrt{t},\beta)$ with $\beta=3.6$, $2.6$ and $2.16$, appear to fit well the numerical data. As to be expected, the fitting parameters $B$ and $D$ in the inset coincide with $B_s=v_0^2/2$ and $D_s=v_0^2/2D_\theta$. (b) $\Delta x$ pdf for increasing $t$ values and
$\bar \phi>\bar \phi_*$ (see legend); the Gaussian ($\beta=2$, solid) and Laplace distributions ($\beta=1$, dashed) for the $D$ value fitting the corresponding data in Fig. \ref{F1}(a), are drawn for reference.
In (c) and (d), $\Delta x$ pdf's are displayed for suspension configurations with the same $\bar \phi$, but resting respectively on the upper ($\beta>2$) and lower branch ($\beta<2$) of the hysteresis loop of Fig.~\ref{F3}(a).
Other simulation parameters are: $\tau_\theta=100$, $v_0=1$, and $N=9,120$. The estimated NGND transient time is $\tau_{\rm NGND}\sim 7.5 \cdot 10^{5}$ (see text).
\label{F4}}
\end{figure}

\section{non-Gaussian normal diffusion and phase separation} So far we have characterized MIPS in
terms of particle self-diffusion under the condition of normal diffusion,
i.e., for $t\gg \tau_\theta$. The question now rises whether this criterion
suffices to define the relaxational properties of the suspension
steady-state. To address this issue we computed the particle displacement
distributions, $P(\Delta x, t)$, at increasing time intervals, $t$. Examples
are reported in Fig. \ref{F4}, for non-interacting particles [panel (a)], interacting particles with $\bar \phi<\bar \phi_*$ [ panels (b) and and (c)] and $\bar \phi>\bar \phi_*$ [panels (d)]. In all cases, the
displacement pdf's keep changing over time even in the normal diffusion
regime of Fig. \ref{F1}(a), which implies that the particle dynamics in large
steady-state active suspensions involves long transients, largely ignored in
previous investigations. Stationary Gaussian distributions were obtained,
indeed, but only for exceedingly long simulation runs. More remarkably, when
plotted versus $\Delta x$ in the normal diffusion regime, the transient
$\Delta x$ pdf's are platykurtic in the homogeneous phase and leptokurtic
under phase separation. With this notation we mean that in the presence
(absence) of phase separation, the $\Delta x$ pdf's have thinner (fatter)
tails than the corresponding asymptotic Gaussian distributions.
It should be remarked that the transition from platy- to lepto-kurtic
transients is not much a signature of the gaseous spinodal crossing,
$\phi^{\rm (s)}_g$, as a property of the cluster phase
itself. Indeed, it has been observed all along the lower (MIPS) branch
of the hysteresis loop for $\phi_g < \bar \phi < \phi^{\rm (s)}_g$ [compare
Figs. \ref{F4}(b,c)].

\begin{figure}[tp]
\centering 
\includegraphics[width=8.5cm]{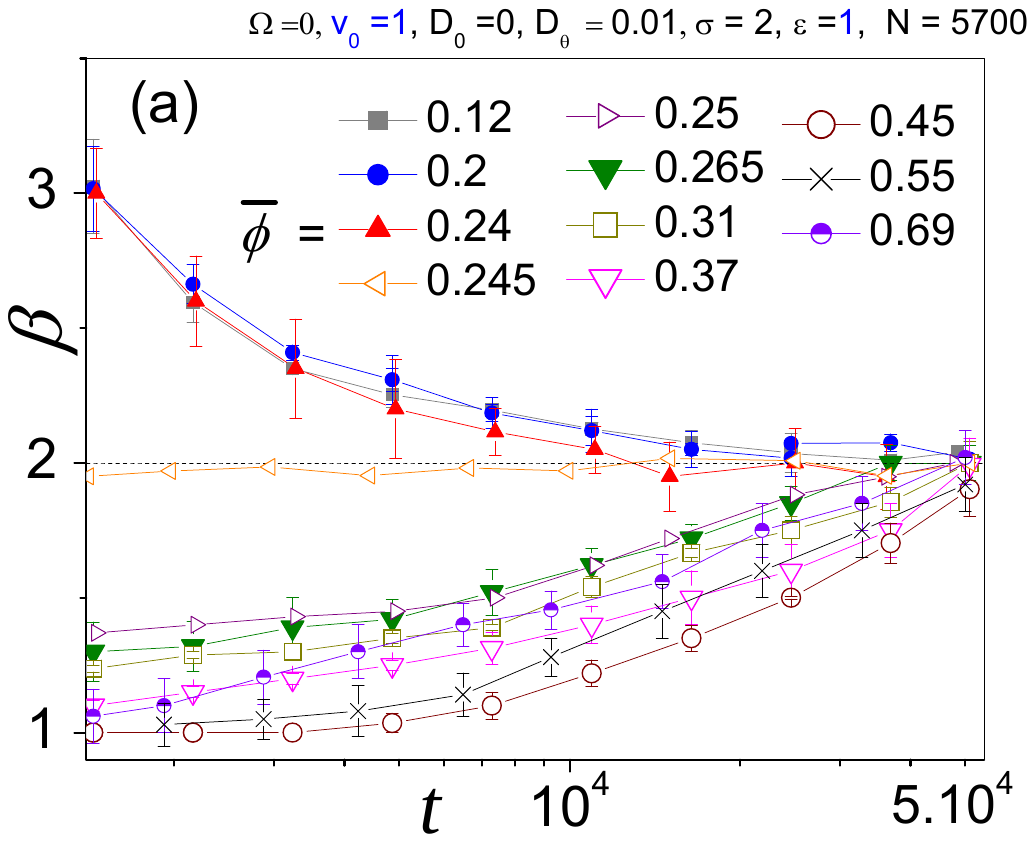}
\includegraphics[width=8.5cm]{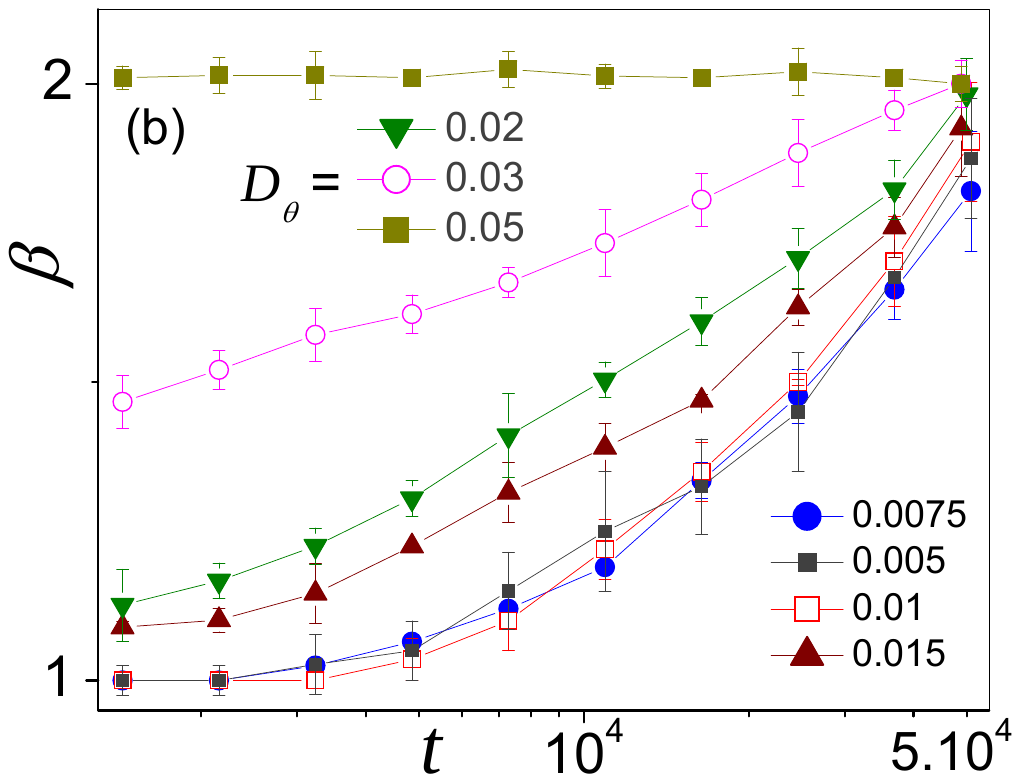}
\caption{Non-Gaussian normal diffusion as a MIPS signature. (a) NGND parameter $\beta$ vs. $t$ for increasing values of $\bar \phi$. Phase coexistence is clearly characterized by leptokurtic transient $P(\Delta x,t)$ with fat tails ($\beta < 2$). (b) $\beta$ vs. $t$ with $\bar \phi=0.45$ and increasing $D_\theta$ (see legends). All simulations were performed assuming a randomly uniform initial particle distribution. Other simulation parameters are: $\tau_\theta=100$, $v_0=1$, and $N=9,120$. The estimated NGND transient time is $\tau_{\rm NGND}\sim 7.5 \cdot 10^{5}$ (see text).
\label{F5}}
\end{figure}

\subsection{Heuristic fitting function bridging Gaussian and non-Gaussian displacement distribution} 
For a more quantitative analysis of the transient $\Delta x$ distributions,
we had recourse to the heuristic fitting function, $P(\Delta x/\sqrt{t},
\beta)$, first introduced in Ref. \cite{FoP} to bridge Gauss ($\beta=2$) and
Laplace (i.e., exponential, $\beta=1$) pdf's with same MSD, $2Dt$, namely,
\begin{eqnarray}
P_G(\Delta x,t; D)&=&\frac{1}{\sqrt{4\pi Dt}} \exp\left[-\frac{\Delta x^2}{4Dt}\right],\;\;
{\rm and} \nonumber \\
 P_L(\Delta x, t; D)&=&\frac{1}{\sqrt{Dt}} \exp\left[-\frac{\Delta x}{\sqrt{Dt}}\right]\nonumber.
\end{eqnarray}
Contrary to the diffusing diffusivity model \cite{Slater}, where the limiting Laplace and
Gaussian distributions are functions of the sole diffusion constant, $D$, a more realistic fitting procedure 
needs at least one additional $t$-dependent parameter, $\beta$, to capture the transient character of the displacement pdf's.
To this purpose, in Ref. \cite{FoP}, we started from the compressed exponential function
\begin{equation} \label{B1}
p(\delta_t)=p_0 \exp{\left[-(\delta_t/\delta_0)^\beta\right]},
\end{equation}
where $\delta_t=\Delta x/\sqrt{t}$ and $\beta \geq 1$. The scaling factor, $\delta_0$, and the
normalization constant, $p_0$, were computed by imposing the conditions
\begin{equation} \label{B2}
\int_0^\infty p(\delta_t)d\delta_t=1, ~~~\int_0^\infty \delta_t^2p(\delta_t)d\delta_t=2D,
\end{equation}
to obtain the one-parameter {\em ad hoc} fitting function,
\begin{equation} \label{B3}
P(\delta_t,\beta)= \frac{\beta}
{\Gamma(\frac{1}{\beta})^\frac{3}{2}} \left [\frac{\Gamma(\frac{3}{\beta})}{2D} \right ]^\frac{1}{2}
\exp \left[-\left(\frac{\delta_t^2}{2D}
\frac{\Gamma(\frac{3}{\beta})}
{\Gamma(\frac{1}{\beta})}
\right)^{\frac{\beta}{2}} \right].
\end{equation}
This is the definition of the fitting functions, $P(\Delta x/\sqrt{t},\beta)$,
plotted in Fig.~\ref{F4}. The fitting parameter $\beta$ is allowed to vary
with $t$; it assumes values in the range $1 \leq \beta \leq 2$ for
leptokurtic distributions (positive excess kurtosis) and $\beta \geq 2$ for
platykurtic distributions (negative excess kurtosis).

The $\beta$ values displayed in Fig.~\ref{F5} have been
generated from Eq. (\ref{B3}) by setting $D$ equal to the diffusion constants
that best fitted the large-$t$ diffusion data in Fig.~\ref{F1}(b) and then computing $\beta$ to get the best fit of the $\Delta
x/\sqrt{t}$ distributions at different $t$. The numerical transients of a
suspension of non-interacting active particles for $t\gtrsim \tau_\theta$,
displayed here in Fig. \ref{F4}(a), are  well reproduced by this
fitting function. The $P(\Delta x/\sqrt{t},\beta)$ spikes at short times, $t
< \tau_\theta$, are centered around $v_0\sqrt{t}$, as to be expected
in the ballistic regime.

We recall that when taking into account steric effects, all distributions
$P(\Delta x, t)$ were computed after the active suspension had reached
its stationary state. This means that for $\bar \phi >\bar \phi_*$, we
started counting $t$ only after MIPS had occurred.  Leptokurtic transients with $t$-dependent $\beta$ are a 
defining MIPS property as proven by the fact that NGND was observed 
along the entire lower branch of the hysteresis loops of Fig.~\ref{F3}(a). 
An example is illustrated in Figs.~\ref{F4} (b) and (c).

 The transient character of the $\Delta x$ distributions
was thus quantified by the $t$-dependent fitting parameter $\beta$. In Fig.
\ref{F5}(a-b) we display $\beta$ vs. $t$ for $v_0=1$, randomly uniform
initial conditions and increasing values of
$\bar \phi$. All curves approach the horizontal asymptote, $\beta=2$, as
expected; more importantly, they do so from above for $\bar \phi <\bar
\phi_*$ and from below for $\bar \phi>\bar \phi_*$. The transition from
platy- to lepto-kurtic transient pdf's is the sharpest at short
normal-diffusion times. This property provides an alternative, but consistent
signature of the MIPS threshold ($\bar \phi_*=0.245\pm 0.005$ for $v_0=1$).
It should be remarked that the transition from platy- to lepto-kurtic
transients is not much a signature of the gaseous phase
spinodal, $\bar \phi=\phi^{\rm (s)}_g$, as a property of the cluster phase
itself. Indeed, it has been observed also along the lower (MIPS) branch
of the hysteresis loop for $\phi_g < \bar \phi < \phi^{\rm (s)}_g$.

The transient displacement distributions, $P(\Delta x,t)$, in a
low-density active suspension are governed solely by diffusion in a
homogeneous phase, whence their platykurtic character \cite{FoP}. Vice versa,
under phase separation the JP's tend to cluster in compact structures made of
hexatic domains of different sizes. Due to simultaneous multiple collisions,
the instantaneous particle diffusion in such aggregates is Gaussian,
with diffusion constant in the range $[0, D_s]$.
As discussed in Ref. \cite{FoP}, correlations of the instantaneous diffusion constant on a
suitably long time-scale suffice to produce leptokurtic profiles of $P(\Delta
x,t)$ with $1<\beta<2$. Here, such correlations are related to the ultra-slow
particle diffusion in the dense phase \cite{Redner1}.  

\begin{figure}[tp]
\centering 
\includegraphics[width=8.5cm]{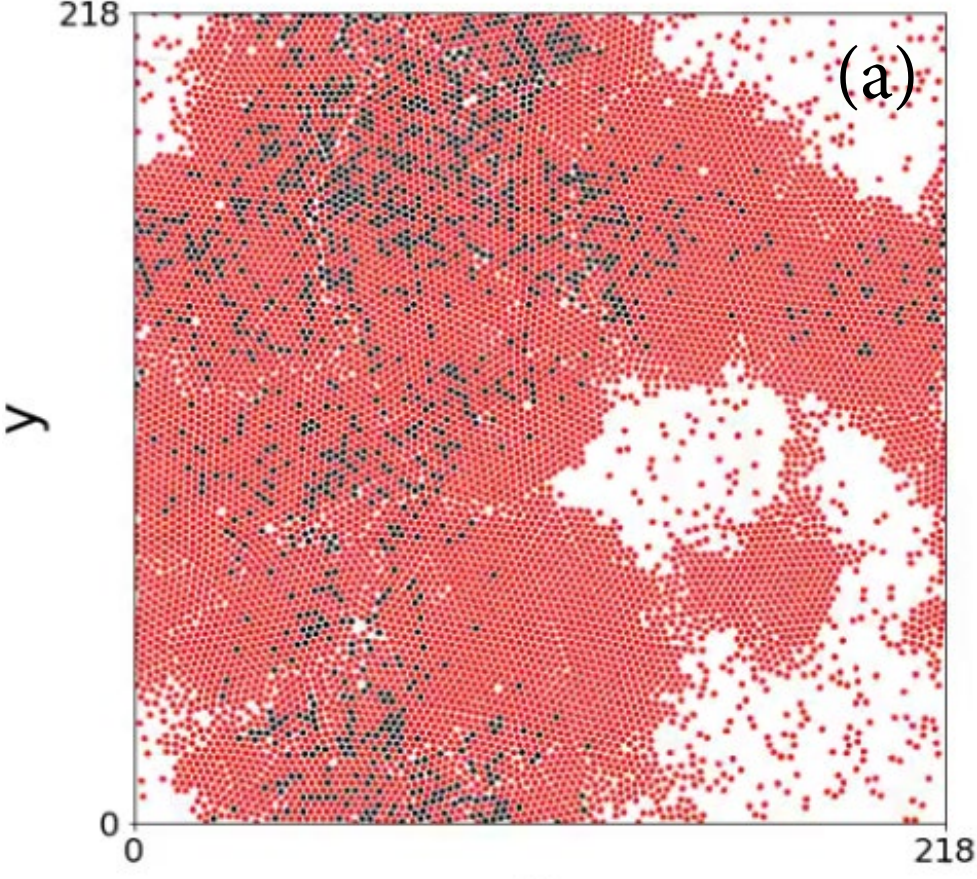}
\includegraphics[width=8.5cm]{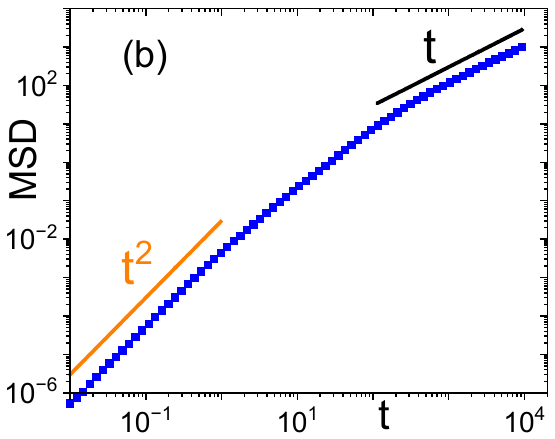}
\caption{In-cluster diffusion in a suspension of $N=9,120$ active JP's with $\bar \phi=0.60$. All remaining simulation parameters are as in Fig. \ref{F1}. (a) Randomly selected particles (black dots) inside the steady-state dense phase at $t=10^4$; (b) MSD vs. $t$ of the selected particles diffusing along the $x$ axis without leaving the dense phase.  The corresponding asymptotic in-cluster diffusion constant, $D_c \sim 10^{-3}D_s$.
\label{F6}}
\end{figure}
\subsection{In-cluster diffusion constant and non-Gaussian normal diffusion transient time} The NGND mechanism
assumes that the suspension has reached a (quasi)stationary state characterized by a
normal average MSD. In view of the ergodic property, %illustrated [see Fig.1 of supporting information~\cite{SM}], 
the particle diffusivity was averaged over the entire suspension, the corresponding self-diffusion constant, $D$ of Fig.~\ref{F1}(b), making no difference between diffusion in the dilute and in the dense phase.
The structure of the dense phase, in particular, is far from homogeneous
\cite{Redner1} and keeps varying slowly with time. As a consequence, the
diffusion of a single particle may appear normal over relatively short time
intervals, but is in fact time modulated on longer time scales. To this
purpose, we numerically estimated
the asymptotic in-cluster diffusion constant, $D_c$, by monitoring the spatial
diffusion of a number of particles randomly selected inside a steady-state
cluster with $\bar \phi>\bar \phi_*$. Only trajectories of particles
not leaving the cluster during the entire simulation run were averaged to
compute the in-cluster MSD versus $t$ (an example is displayed in Fig.~\ref{F6}). The resulting diffusion constant, $D_c \sim 10^{-3}D_s$, weakly depends
on $\bar \phi$ and defines a characteristic NGND transient time, $\tau_{\rm NGND}\sim \alpha_c L^2/8D_c$,
($\alpha_c^{{1}/{2}} L$ being the average cluster diameter), in good agreement
with the numerical data of Fig.~\ref{F5}.

\section{Conclusion} We characterized MIPS of an athermal, achiral active suspension by
looking at the particle diffusivity under steady-state conditions. The choice
of using the overall suspension packing fraction as tunable parameter has a
practical motivation as in most applications the particle motility cannot be
varied at will, whilst their density can. Particle diffusion under phase separation has been
proven to show hysteretic and NGND properties. Our main conclusions are:
(1) The hysteresis loop of the curve $D(\phi)$ in the lower binodal region which allows a direct measure of  $\phi_g$  and $\phi_g^{(s)}$.
(2) The peculiar properties of the upper binodal and spinodal curves, which appear to overlap, thus suppressing hysteresis in the upper binodal region.  Our numerical data also suggest a mirror symmetry of the spinodal curves with  $\phi_g^{(s)}+\phi_c^{(s)}$ =1.
(3) Non-Gaussian normal diffusion  characterized motility induced phase separation with leptokurtic transient distributions of displacements $\Delta x$. The associated NGND transient time is almost four orders of magnitude larger than the rotational relaxation time of a free JP. Further, we show that NGND characterized MIPS also in the presence of hysteresis.

\section*{Acknowledgements}
Y.L. is supported by the NSF China under grant No. 12375037, and F.M.
also by NSF China under grant No. 12350710786. P.K.G. is supported by
SERB Core Research Grant No. CRG/2021/007394. P.B. thanks UGC,
New Delhi, India, for the award of a Junior Research Fellowship.
We thank RIKEN supercomputer Hokusai for providing computational resources. F.N. is supported in part by: the Japan Science and Technology Agency (JST)
[via the CREST Quantum Frontiers program Grant No. JPMJCR24I2,
the Quantum Leap Flagship Program (Q-LEAP), and the Moonshot R \& D Grant Number JPMJMS2061],
and the Office of Naval Research (ONR) Global (via Grant No. N62909-23-1-2074).

\end{document}